# Why In$_2$O$_3$ Can Make 0.7 nm Atomic Layer Thin Transistors?

Mengwei Si[1], Yaoqiao Hu[2], Zehao Lin[1], Xing Sun[3], Adam Charnas[1], Dongqi Zheng[1], Xiao Lyu[1], Haiyan Wang[3], Kyeongjae Cho[2], and Peide D. Ye[1,*]

[1]*School of Electrical and Computer Engineering and Birck Nanotechnology Center, Purdue University, West Lafayette, IN 47907, United States*

[2]*Department of Materials Science and Engineering, The University of Texas at Dallas, Richardson, Texas 75080, USA*

[3]*School of Materials Engineering, Purdue University, West Lafayette, IN 47907, United States*

* Address correspondence to: yep@purdue.edu (P.D.Y.)

ABSTRACT

In this work, we demonstrate enhancement-mode field-effect transistors by atomic-layer-deposited (ALD) amorphous $In_2O_3$ channel with thickness down to 0.7 nm. Thickness is found to be critical on the materials and electron transport of $In_2O_3$. Controllable thickness of $In_2O_3$ at atomic scale enables the design of sufficient 2D carrier density in the $In_2O_3$ channel integrated with the conventional dielectric. The threshold voltage and channel carrier density are found to be considerably tuned by channel thickness. Such phenomenon is understood by the trap neutral level (TNL) model where the Fermi-level tends to align deeply inside the conduction band of $In_2O_3$ and can be modulated to the bandgap in atomic layer thin $In_2O_3$ due to quantum confinement effect, which is confirmed by density function theory (DFT) calculation. The demonstration of enhancement-mode amorphous $In_2O_3$ transistors suggests $In_2O_3$ is a competitive channel material for back-end-of-line (BEOL) compatible transistors and monolithic 3D integration applications.

KEYWORDS: indium oxide, oxide semiconductor, thin-film transistor, charge neutrality level, ultrathin body, enhancement-mode

Amorphous oxide semiconductors, such as Indium-Gallium-Zinc-Oxide (IGZO), are leading channel materials in thin-film transistors (TFTs) for flat-panel display applications.[1] Recently, oxide semiconductors, such as indium oxide ($In_2O_3$), indium-tin-oxide (ITO)[2-3], W-doped $In_2O_3$ (IWO)[4], IGZO[5,6], attracted revived interest because it can be applied in back-end-of-line (BEOL) compatible transistors for monolithic 3D integration. Although $In_2O_3$ has a much higher electron mobility than quaternary IGZO, IGZO was adopted and commercialized in TFT technology rather than $In_2O_3$ because of two reasons.[7] First, $In_2O_3$ form poly-crystalline films deposited by sputtering, resulting in unstable electrical properties due to grain boundaries. Second, electron density of bulk $In_2O_3$ films is very high on the order of $10^{20}$ /cm$^3$ and difficult to control so that enhancement-mode operation cannot be achieved to suppress off-current ($I_{OFF}$) at zero gate bias. These two major challenges prevented $In_2O_3$ being employed in TFT technology for display applications and also limit the application of $In_2O_3$ as channel for BEOL compatible transistors.

Here, enhancement-mode amorphous $In_2O_3$ transistors are clearly demonstrated to overcome the above two limitations, by applying atomic layer ultrathin $In_2O_3$ film enabled by atomic layer deposition (ALD). The thickness of $In_2O_3$ is the most critical parameter, varying from 0.7 nm to 1.5 nm here. First, ultrathin $In_2O_3$ causes the reduction of 2D channel carrier density and the realization of enhancement-mode $In_2O_3$ transistors. Second, ALD grown ultrathin $In_2O_3$ film[8–11] is amorphous, confirmed by high resolution transmission electron microscopy (HRTEM). It is also well known that the degree of crystallinity decreases while reducing the film thickness down to 2-3 nm in certain oxides.[12] More importantly, this work unveils the interface physics why $In_2O_3$ can be scaled down to 0.7 nm as thin as monolayer of $MoS_2$, in great contrast to the thickness scaling limit about 3 nm for Si and 10 nm for GaAs.

The amorphous $In_2O_3$ transistors exhibit clear switching characteristics even with only 0.7 nm thick channel, showing on/off ratio over 7 orders and enhancement-mode operation. Carrier transport mechanism in atomically thin 3D semiconducting channel less than 1 nm is *unexplored*. In this work, carrier transport is investigated on $In_2O_3$ transistors with various channel thickness ($T_{ch}$) from 1.5 nm down to 0.7 nm. The threshold voltage ($V_T$) of $In_2O_3$ transistors are found to be considerably tuned by $T_{ch}$ from depletion-mode to enhancement-mode. It is understood that trap neutral level (TNL) alignment[13-24] to ultrathin $In_2O_3$ plays a critical role in all these $In_2O_3$ transistor characteristics. In bulk or thick $In_2O_3$ films, TNL lies far above conduction band minimum ($E_C$)[24], so that un-doped bulk $In_2O_3$ film is always n-type with high carrier density, as shown in Fig. S1 in supporting information. Depletion-mode $In_2O_3$ transistors with large drain currents are easy to be demonstrated due to high electron density and mobility.[3] In ultrathin $In_2O_3$ film down to nanometer scale, the band gap is strongly affected by quantum confinement effect like 2D van der Waals materials. Therefore, TNL can be turned from far above $E_C$ to below $E_C$, by decreasing $T_{ch}$ from 1.5 nm to 0.7 nm, due to the bandgap enhancement. Therefore, enhancement-mode operation can be achieved on $In_2O_3$ transistors although the drain currents are dropped due to the reduction of carrier density and mobility.

The demonstration of enhancement-mode amorphous $In_2O_3$ transistor suggests $In_2O_3$ is a competitive channel material for TFT technology in display applications and for BEOL compatible transistor applications. From device application point of view, it offers general advantages over 2D van der Waals materials such as wafer size homogeneous deposition, ALD dielectric integration, and ultimate scaling due to atomic layer thin channel and relatively low dielectric constant of 8.9 for $In_2O_3$.[25] Moreover, ALD $In_2O_3$ enables tremendous new opportunities for BEOL device process and integration because of its BEOL compatible low temperature process, wafer-scale homogenous films, atomically thin and smooth surface for ultimately scaled devices, high mobility, and more importantly, the conformality on side walls, deep trenches, 3D structures for 3D device integration.

Fig. 1(a) shows the schematic diagram of a back-gate $In_2O_3$ TFT fabricated on silicon substrate. Fig. 1(b) shows the gate stack with detailed layer thicknesses. The gate stack consists of 40 nm W as gate metal, 10 nm $HfO_2$ and 1 nm $Al_2O_3$ as gate dielectrics, 0.7-1.5 nm $In_2O_3$ as semiconducting channels and 80 nm Ni as source/drain (S/D) ohmic contacts. 1 nm $Al_2O_3$ is applied to protect $HfO_2$ during processing and improve the interface quality. The device fabrication process is described in methods section and has a low thermal budge of 225 °C as BEOL compatible device technology. Fig. 1(c) presents a scanning electron microscopy (SEM) image of a fabricated $In_2O_3$ transistor, capturing the W gate metal and Ni S/D ohmic contacts. Note that the nanometer-thin $In_2O_3$ channel is too thin to be visible under SEM, the thickness of which is determined together by atomic force microscopy (AFM) and HRTEM. The inset of Fig. 1(c) is the AFM measurement of $In_2O_3$ on a fabricated $In_2O_3$ with a measured $T_{ch}$ of 0.7 nm. Cross-sectional HRTEM images of 0.7 nm and 1.2 nm thick $In_2O_3$ are illustrated in Fig. 1(d) and 1(e), respectively. $In_2O_3$ transistors with various $T_{ch}$ of 0.7, 1, 1.2, 1.5 nm are fabricated and studied. The HRTEM images demonstrate that ultrathin $In_2O_3$ films in nanometer scale by ALD is amorphous, which is very different from bulk poly-crystalline $In_2O_3$ films deposited by sputtering[7]. This amorphous property most likely originates from the thickness-dependent crystallinity[12], which is widely reported in various oxides.

Fig. 2(a) shows the $I_D$-$V_{GS}$ characteristics of an $In_2O_3$ transistor with channel length ($L_{ch}$) of 0.2 μm and $T_{ch}$ of 0.7 nm, exhibiting on/off ratio over 7 orders and $V_T$ of 4.9 V extracted by linear extrapolation. The In$_2$O$_3$ channel is composed of only a few atoms vertically. In$_2$O$_3$ also has low dielectric constant of 8.9, which further enhances the gate electrostatic control. The ultrathin channel and low dielectric constant properties can improve the immunity to short channel effects for ultra-scaled BEOL logic application. The I$_D$-V$_{DS}$ characteristics of an In$_2$O$_3$ transistor with L$_{ch}$ of 0.2 μm and T$_{ch}$ of 0.7 nm are presented in Fig. 2(b), with a maximum drain current of 0.5 μA/μm at V$_{DS}$=4 V. Fig. 2(c) shows the I$_D$-V$_{GS}$ characteristics of an In$_2$O$_3$ transistor with L$_{ch}$ of 0.2 μm and T$_{ch}$ of 1 nm, exhibiting on/off ratio over 9 orders and V$_T$ of 3.0 V. Enhancement mode operation is also achieved with 1-nm thick In$_2$O$_3$ channel. The corresponding I$_D$-V$_{DS}$ curve is shown in Fig. 2(d), with a maximum drain current improved to 96 μA/μm at V$_{DS}$=3 V. Fig. 2(e) illustrates the I$_D$-V$_{GS}$ characteristics of an In$_2$O$_3$ transistor with L$_{ch}$ of 0.2 μm and T$_{ch}$ of 1.2 nm, exhibiting on/off ratio over 10 orders of magnitude and V$_T$ of 0.3 V. The I$_D$-V$_{DS}$ characteristics are shown in Fig. 2(f), with a maximum drain current of 503 μA/μm at V$_{DS}$=1.2 V. Fig. 2(g) presents the I$_D$-V$_{GS}$ characteristics of an In$_2$O$_3$ transistor with L$_{ch}$ of 0.2 μm and T$_{ch}$ of 1.5 nm, exhibiting on/off ratio over 7 orders of magnitude and V$_T$ of -3.8 V. The I$_D$-V$_{DS}$ characteristics is shown in Fig. 2(h), with a maximum drain current of 1 mA/μm at V$_{DS}$=1 V. All these device performances are among the best for thin-film transistor technology, not counting its atomic thin channel and the potential enhancement by further scaling.

As can be seen, the electron transport in ultrathin In$_2$O$_3$ is very different from that of bulk In$_2$O$_3$. Thickness-dependent electron transport properties are studied statistically in Fig. 3, where each data point represents the average of at least 5 devices with error bar as standard deviation. The very small standard deviation confirms the highly uniformity of ALD grown In$_2$O$_3$. Fig. 3(a) shows the thickness-dependent V$_T$ of In$_2$O$_3$ transistors with L$_{ch}$=0.2 μm. V$_T$ changes from 4.5 V to -3.8 V by increasing T$_{ch}$ from 0.7 nm to 1.5 nm. Both enhancement-mode and depletion-mode are achieved by $T_{ch}$ control. The thickness-dependent field-effective mobility ($\mu_{FE}$) is shown in Fig. 3(b), where $\mu_{FE}$ is extracted from transconductance ($g_m$) at low $V_{DS}$ of 0.1 V. $\mu_{FE}$ decreases exponentially with $T_{ch}$, suggesting stronger disorder induced potential fluctuation in $E_C$ and electron scattering in atomic layer thin $In_2O_3$ film. Fig. 3(c) presents the $g_m$ *versus* $T_{ch}$ at $V_{DS}$=1 V of $In_2O_3$ transistors with $L_{ch}$ of 0.2 μm. $g_m$ increases with thicker $T_{ch}$ due to a higher mobility and a maximum $g_m$ of 162 μS/μm at $T_{ch}$ of 1.5 nm is achieved at $V_{DS}$=1 V.

Why all of sudden can we scale down 3D semiconductor channel down to atomic layer thin 0.7 nm to 1.5 nm, comparable to the thickness of monolayer or bi-layer of 2D van der Waals materials and far thinner than those limits for conventional 3D semiconductors such as Si and GaAs? Fermi-level pinning concept is extremely important in semiconductor materials and device development and various models were developed spanning several decades.[13-24,26,27] Since the physics and chemistry of interfaces are very complex, there are no comprehensive models for both metal/semiconductor and semiconductor/dielectric interfaces. The charge neutrality level (CNL) model is widely applied to describe metal/semiconductor interface.[16] Meanwhile, we can also use so called trap neutral level (TNL) to describe all the experimental observation in III-V and Ge with ALD dielectrics.[20,21,23] More interestingly, the energy level and its alignments of CNL, TNL or even defect energy level in all bulk semiconductors are eventually strongly correlated and similar. From the perspective of basic physical concepts, such agreement is because all of these phenomena are related to defects, no matter whether these defects are located at interfaces or in the semiconductor bulks.

First of all, CNL in bulk $In_2O_3$ lies about 0.4 eV above $E_C$ so that thick $In_2O_3$ film is considered as conducting oxide[24,26]. The similar band alignment can be found in $SnO_2$ about 0.5 eV above $E_C$ so that ITO as a combination of $In_2O_3$ and $SnO_2$ has an ultra-high electron density[3] and is widely used as transparent conductor. The location of $E_F$ in $In_2O_3$ at semiconductor/dielectric interface is known to be determined by the TNL.[21,23] Similar to CNL, TNL at $In_2O_3/Al_2O_3$ interface would also align above $E_C$ if $In_2O_3$ channel is 1.5 nm or thicker and only depletion-mode operation can be realized in thicker channels. Conventional high-k dielectric can only modulate carrier density up to $2\text{-}3\times10^{13}$ /cm$^2$ so that the channel cannot be depleted if the channel is thicker than 2 nm.[3] Therefore, if not applying gate voltage (approximately similar to $V_{GS}$=0 V assuming flat-band voltage around 0), Fermi level ($E_F$) is above $E_C$ for thick $In_2O_3$ films while $E_F$ is below $E_C$ for thinner $In_2O_3$ films at atomic layer scale. The thickness-dependent $V_T$ presented in Fig. 3(a) clearly shows this trend. TNL in $In_2O_3$ moves deeply below $E_C$ once the channel thickness becomes much thinner than 1 nm.

The control of TNL alignment by thickness control can be understood by the quantum confinement effect like layer dependent band-structures in 2D van der Waals materials. In $In_2O_3$ transistor structure as shown in Fig 1(a), the semiconducting $In_2O_3$ is sandwiched by insulating $Al_2O_3$ and air, so that electron transport in $In_2O_3$ behaves like 2D electron gas in an infinite quantum well. The change of $E_C$ due to the quantum confinement effect is like the ground state energy for the electron in an infinite potential well. To verify this mechanism, DFT modelling is performed to investigate how $E_C$ changes with $In_2O_3$ thickness. The DFT model consists of a corundum type $In_2O_3$ layer (representing amorphous $In_2O_3$ layer) sandwich stacked between two corundum $Al_2O_3$ layers, as shown in Fig. 4(a). The DFT model consists of an $In_2O_3$ layer stacked on an $Al_2O_3$ layer, as shown in Fig. 4(a). For ALD-grown $In_2O_3$ film, its surface would be naturally terminated by -OH group, therefore it is reasonable to terminate the $In_2O_3$ surface with H atoms. The large conduction band offset at $In_2O_3/Al_2O_3$ interface (> 4 eV)[27] guarantees a sufficient high potential barrier to introduce the quantum confinement effect on $In_2O_3$ layers. The calculated thickness-dependent band structures along Γ-X direction are shown in Fig. 4(b). The energy bands of $Al_2O_3$ are set as reference to show the position shift of band edges of $In_2O_3$ thin films. When $In_2O_3$ thickness decreases, $E_C$ moves up in the absolute energy scale, although $E_V$ remains almost unchanged. This clearly manifests the thickness-dependent quantum confinement effect in ultrathin $In_2O_3$ films. Since TNL is the intrinsic property of the material so that it is independent on the channel thickness. For $In_2O_3$ bulk, the TNL is located ~0.4 eV above $E_C$.[24,26] When the $In_2O_3$ film is thinned down to 1.5 nm, $E_C$ upshifts by ~0.6 eV, indicating that the TNL in this case is located ~0.2 eV below $E_C$. As the $In_2O_3$ thickness further decreases, the TNL continuously shifts relatively down toward the mid gap. Therefore, TNL moves deeper inside bandgap while decreasing the $T_{ch}$, resulting in the reduction of carrier density and positive $V_T$ shift, which agrees well with the experimental data in Fig. 2 and Fig. 3. Such quantum confinement effect can also be estimated analytically, as shown in supporting information section 3, which also agrees with the DFT calculation and experimental results.

To switch off $In_2O_3$ transistor with TNL far above $E_C$ or to switch on $In_2O_3$ transistor with TNL deeply below $E_C$, a large amount of trapped charge will be generated due to the assumed U-shape trap density distribution[21], leading to much less effective gate control and not able to demonstrate a well-behaved transistor, as shown in Fig. S2(a) and S2(b). Thus, a proper TNL alignment with an appropriate channel thickness can give both high on-current and enhancement-mode operation, as illustrated in Fig. S2(c). In short, the requirements to obtain an ultrathin semiconducting channel at nanometer scale requires a semiconducting material to be highly conductive in bulk while with a bandgap and a low interface trap density.

The above TNL model focuses on the intrinsic thickness-dependent electronic structure without considering the thickness-dependent material structures due to material growth and device fabrication, such as defects and surface roughness. In real experiments, the defect density in $In_2O_3$ films with different thicknesses may be different, which may also lead to a thickness-dependent transport phenomenon. For example, thinner film may be considered as more exposed to the environment so that with less oxygen vacancies. Therefore, the understanding of experimental device characteristics such as $V_T$ shifts should consider the impact from both quantum confinement effects on TNL alignment and the non-ideal electronic defects.

In conclusion, film thickness is found to be critical on the materials and electron transport properties of 3D semiconducting $In_2O_3$. Ultrathin $In_2O_3$ down to 0.7 nm enabled by ALD overcomes two major challenges in $In_2O_3$ TFT technology, i.e. amorphous phase and too high carrier density in $In_2O_3$ channel. The strong thickness-dependent electron transport is understood by the quantum confinement effect on the alignment of TNL. The calculated quantum confined TNL locations agree well with the experimental data. The demonstration of high-performance enhancement-mode amorphous $In_2O_3$ transistors suggests $In_2O_3$ is a competitive channel material for TFT in display applications and for BEOL compatible transistor applications.

METHODS

**Device Fabrication.** The device fabrication process started with solvent cleaning of p+ Si substrate with dry oxidized 90 nm $SiO_2$. 10 nm $Al_2O_3$ was then deposited by ALD at 175 °C with $(CH_3)_3Al$ (TMA) and $H_2O$ as Al and O precursors. W gate metal was then deposited by sputtering, followed by a $CF_4$/Ar ICP dry etching, using $Al_2O_3$ as high selectivity etch stop layer. 10 nm $HfO_2$ and 1 nm $Al_2O_3$ as gate insulator were deposited by ALD at 200 °C with [(CH$_3$)$_2$N]$_4$Hf (TDMAHf) and H$_2$O as Hf and O precursors. In$_2$O$_3$ thin films with various thicknesses were deposited by ALD at 225 °C using (CH$_3$)$_3$In (TMIn) and H$_2$O as In and O precursors. Concentrated hydrochloric acid was employed for the channel isolation. S/D ohmic contacts were formed by e-beam evaporation of 80 nm Ni.

**Device Characterization.** The thickness of the In$_2$O$_3$ was determined together by AFM and TEM. AFM measurement was done with a Veeco Dimension 3100 atomic force microscope system. FEI TALOS F200X operated at 200 kV equipped with super-X electron-dispersive X-ray spectroscopy was used for TEM imaging. The TEM samples were prepared by conventional TEM sample preparation method involving mechanical thinning, polishing and final ion polishing steps. SEM imaging was performed with a Thermo Scientific Apreo S scanning electron microscope. Electrical characterization was carried out with a Keysight B1500 system and with a Cascade Summit probe station in dark and N$_2$ environments at room temperature and at atmosphere. The probe station has a closed chamber to protect the devices from O$_2$ and water in the environments.

**DFT Calculation.** The quantum confinement effect on In$_2$O$_3$ thin film was theoretically characterized by DFT as implemented in Vienna ab initio simulation package (VASP)[28,29] using projected augmented wave (PAW)[30,31]. A model consisting of vacuum/In$_2$O$_3$/Al$_2$O$_3$ layers was used to investigate the quantum confinement effect in ultrathin In$_2$O$_3$. The Perdew-Burke-Ernzerhof generalized gradient approximation (GGA-PBE) functional[32,33] was used to describe the exchange and correlation interaction. A cut-off energy of 420 eV was used for all the calculations. The converged energy criterion for structure relaxation is the force exerted on each atom less than 0.01 eVÅ$^{-1}$. The converged energy criterion for electronic minimization is $10^{-5}$ eV.

ASSOCIATED CONTENT

**Supporting Information**

The supporting information is available free of charge on the ACS Publication website. Additional details for TLM measurements on thick $In_2O_3$ film, the impact of TNL alignments and an analytical quantum confinement model are in the supporting information.

The authors declare no competing financial interest.

AUTHOR INFORMATION

**Corresponding Author**

*E-mail: yep@purdue.edu

**Author Contributions**

P.D.Y. conceived the idea and supervised experiments. M.S. deposited $In_2O_3$ film by atomic layer deposition. X.L. did the W sputtering. M.S., Z.L., A.C. and D.Z. performed device fabrication. M.S. and Z.L. did the electrical measurements. M.S. and P.D.Y. analyzed the electrical data. X.S. and H.W. conducted the TEM characterization. Y.H. and K.C. did the DFT calculation. P.D.Y. and M.S. calculated the analytical infinite quantum well TNL model on ultrathin $In_2O_3$. All authors co-wrote the manuscript.

ACKNOWLEDGEMENTS

The work is mainly supported by SRC nCore IMPACT Center. The work is also partly supported by AFOSR and SRC/DARPA JUMP ASCENT Center. X. S. and H.W. acknowledge the support from the U.S. National Science Foundation for the TEM work (DMR-2016453).

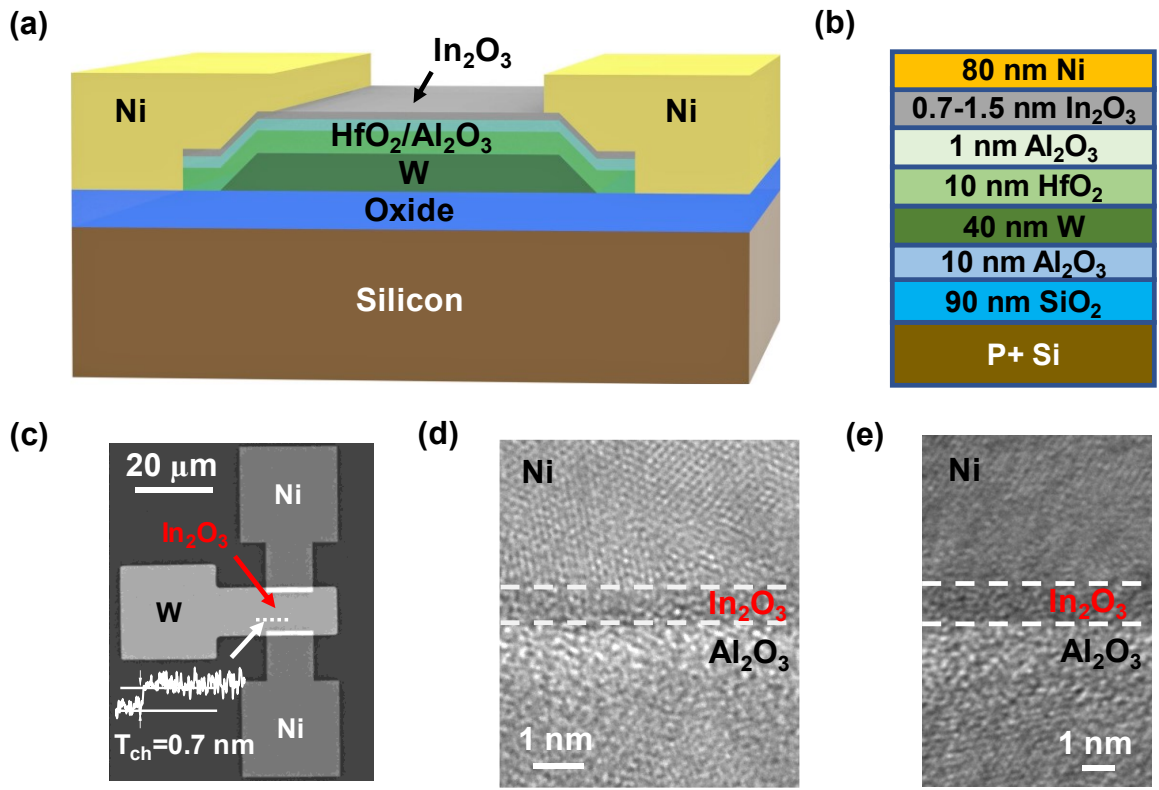

**Figure 1.** (a) Schematic diagram of an $In_2O_3$ transistor. (b) Gate stack of $In_2O_3$ transistors. 40 nm W is used as gate metal, 10 nm $HfO_2$/1 nm $Al_2O_3$ bi-layer is used as gate insulator and 80 nm Ni is used for source/drain electrodes. (c) SEM and AFM measurements of $In_2O_3$ transistor with channel thickness of 0.7 nm. Cross-sectional HRTEM images of (d) 0.7 nm and (e) 1.2 nm thick $In_2O_3$, confirming the thickness of $In_2O_3$ and the amorphous phase.

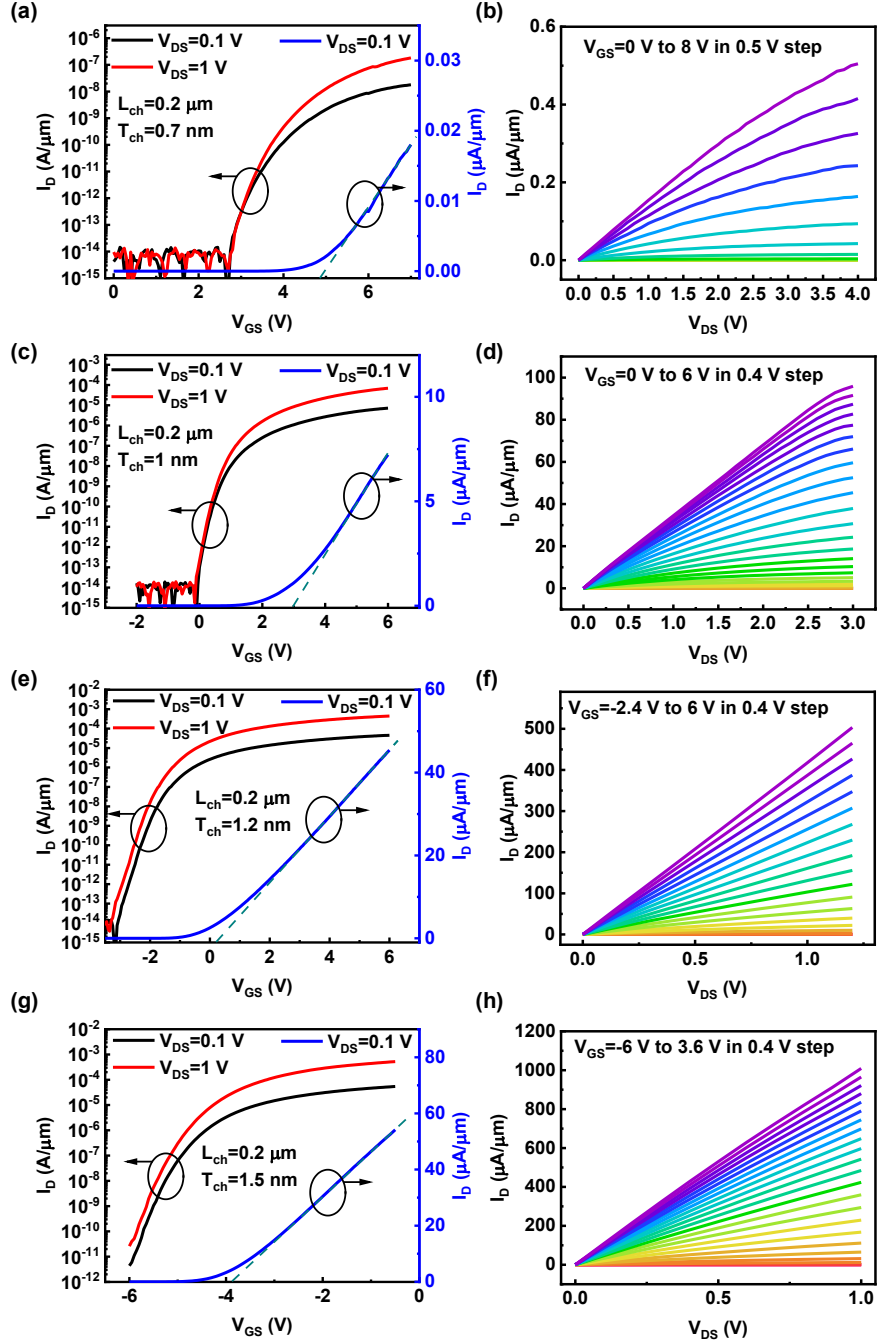

**Figure 2.** (a) $I_D$-$V_{GS}$ and (b) $I_D$-$V_{DS}$ characteristics of an $In_2O_3$ transistor with channel length of 0.2 μm and channel thickness of 0.7 nm, exhibiting on/off ratio > $10^7$ and enhancement-mode operation. (c) $I_D$-$V_{GS}$ and (d) $I_D$-$V_{DS}$ characteristics of an $In_2O_3$ transistor with channel length of 0.2 μm and channel thickness of 1 nm. (e) $I_D$-$V_{GS}$ and (f) $I_D$-$V_{DS}$ characteristics of an $In_2O_3$ transistor with channel length of 0.2 μm and channel thickness of 1.2 nm. (g) $I_D$-$V_{GS}$ and (h) $I_D$-$V_{DS}$ characteristics of an $In_2O_3$ transistor with channel length of 0.2 μm and channel thickness of 1.2 nm, showing on-current > 1 A/mm.

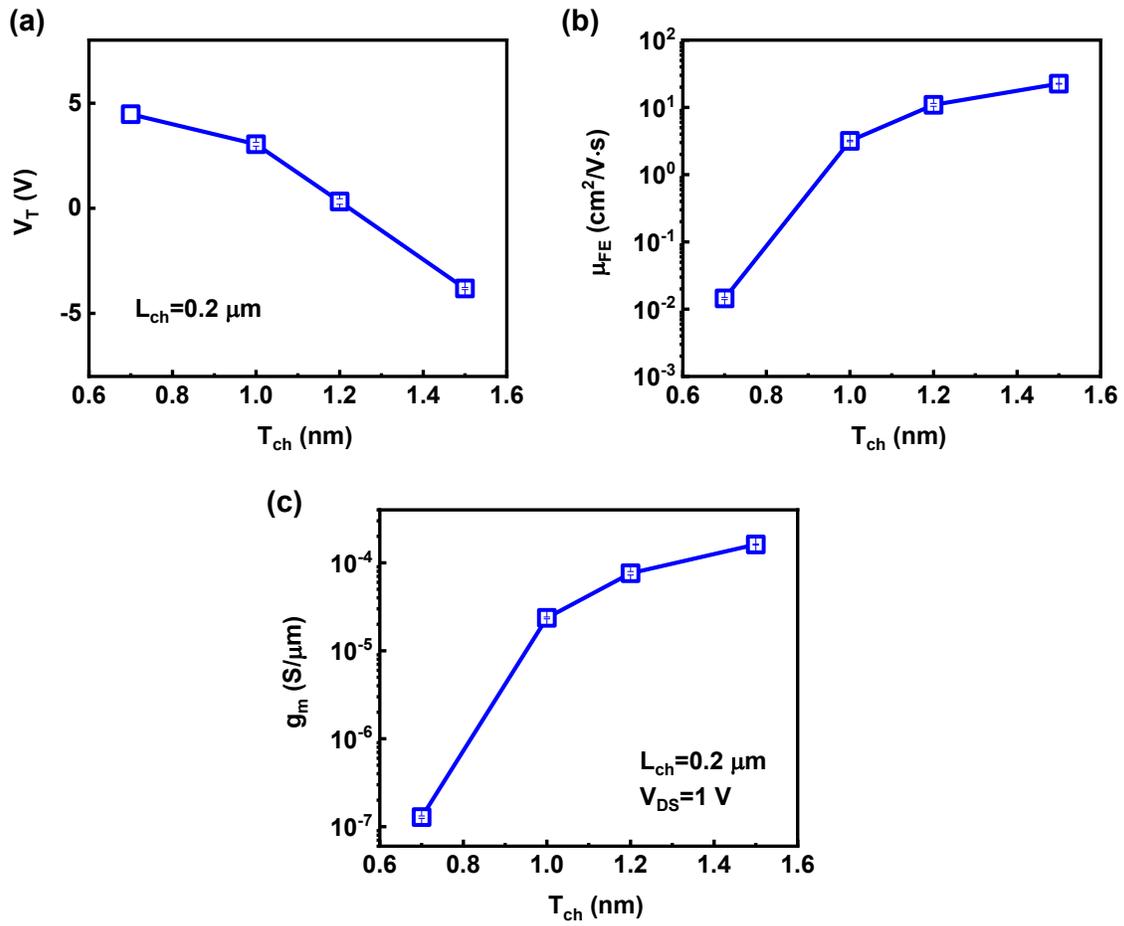

**Figure 3.** (a) Threshold voltage, (b) field effective mobility and (c) transconductance at $V_{DS}=1$ V *versus* channel thickness of $In_2O_3$ MOSFETs with channel length of 0.2 μm. Each data point represents the average of at least 5 devices.

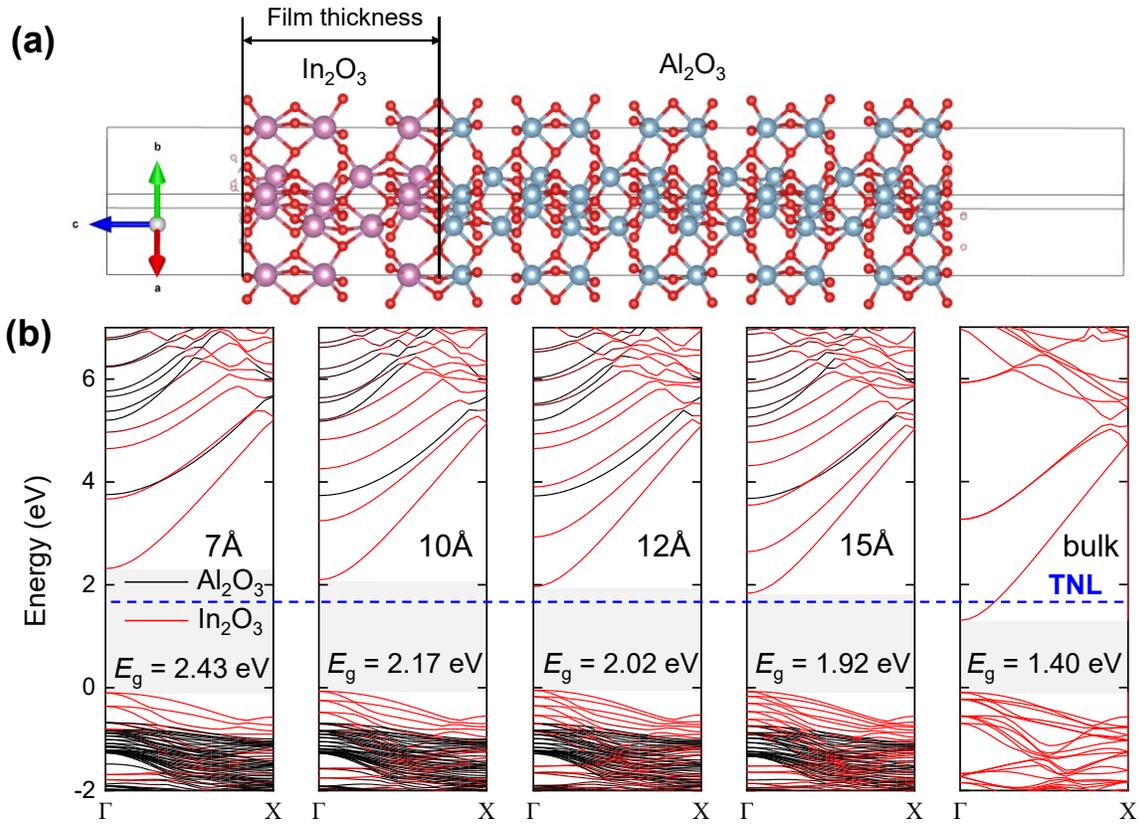

**Figure 4.** (a) Atomic structure of In$_2$O$_3$ film used in the DFT calculations. The film thickness of In$_2$O$_3$ layer is varied to simulate the quantum confinement effect. (b) Calculated band structure at In$_2$O$_3$ film thicknesses of 0.7 nm, 1.0 nm, 1.2 nm, 1.5 nm and bulk along the Γ to X-direction. The red and black lines represent the bands contributed from In$_2$O$_3$ and Al$_2$O$_3$, respectively. The shadowed areas indicate the band gaps of In$_2$O$_3$. Horizontal black dashed lines outline the band edges of Al$_2$O$_3$. TNL is indicated at 0.4 eV above the E$_C$ of bulk In$_2$O$_3$, where TNL alignments relative to conduction band minima are modulated by the thickness of In$_2$O$_3$ due to quantum confinement.

TOC

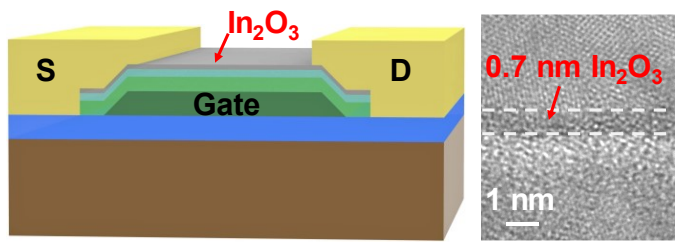

Supporting Information for:

# Why In$_2$O$_3$ Can Make 0.7 nm Atomic Layer Thin Transistors?

Mengwei Si[1], Yaoqiao Hu[2], Zehao Lin[1], Xing Sun[3], Adam Charnas[1], Dongqi Zheng[1], Xiao Lyu[1], Haiyan Wang[3], Kyeongjae Cho[2], and Peide D. Ye[1,*]

[1]School of Electrical and Computer Engineering and Birck Nanotechnology Center, Purdue University, West Lafayette, IN 47907, United States

[2]Department of Materials Science and Engineering, The University of Texas at Dallas, Richardson, Texas 75080, USA

[3]School of Materials Engineering, Purdue University, West Lafayette, IN 47907, United States

* Address correspondence to: yep@purdue.edu (P.D.Y.)

## 1. TLM Measurements on Thick ALD $In_2O_3$

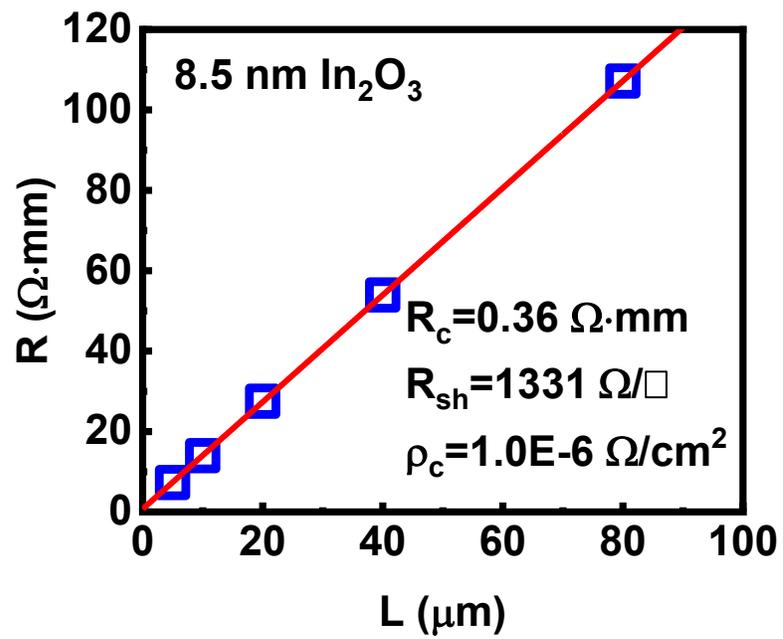

**Figure S1.** TLM measurement of 8.5 nm $In_2O_3$ on 90 nm $SiO_2$/Si substrate with 80 nm Ni contacts, exhibiting $R_C$ of 0.36 Ω·mm, $R_{sh}$ of 1333 Ω/□ and $\rho_c$ of 1.0×10$^{-6}$ Ω/cm².

## 2. The Impact of TNL Alignments

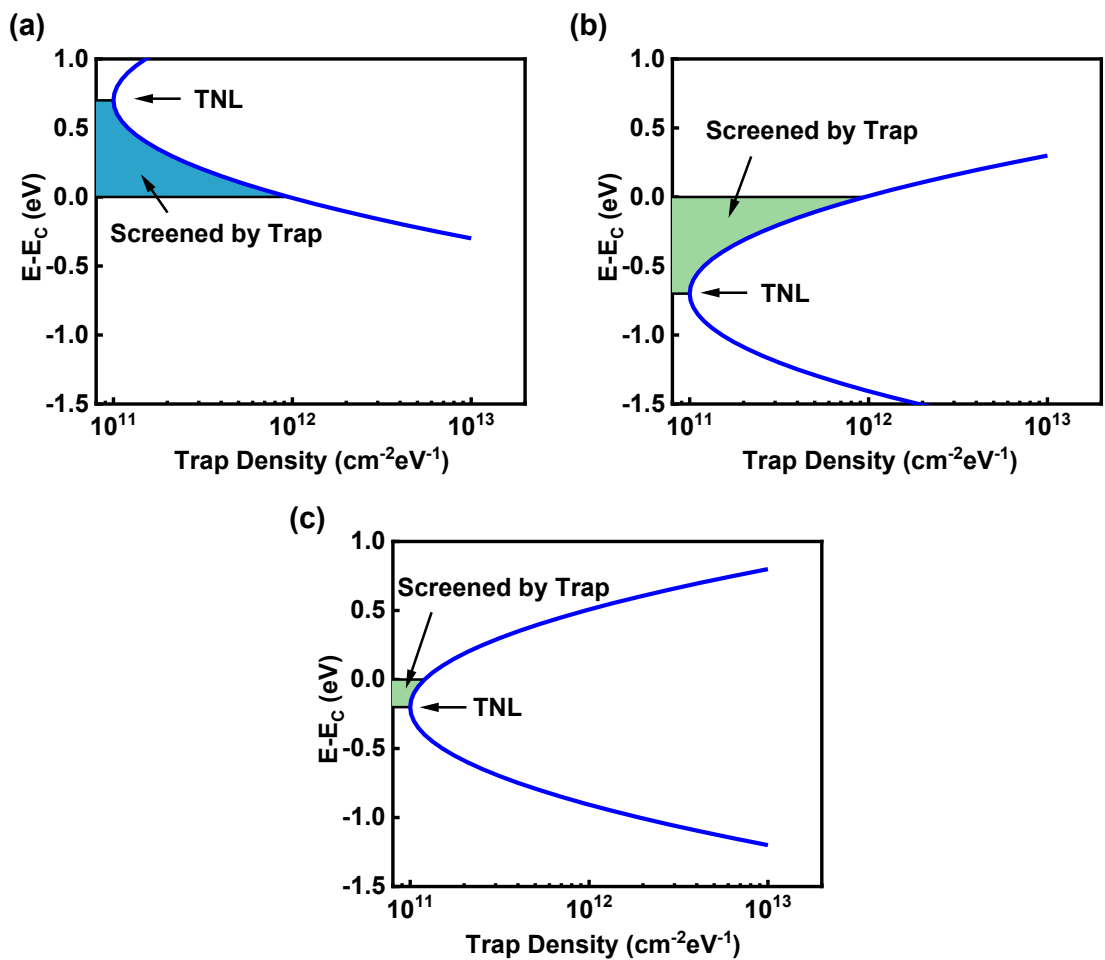

**Figure S2.** Energy level with respect to $E_C$ *versus* trap density for $In_2O_3$ with different TNL locations at (a) TNL far above $E_C$, (b) TNL deeply below $E_C$ and (c) TNL near $E_C$.

## 3. Infinity Quantum Well Model on TNL Alignments

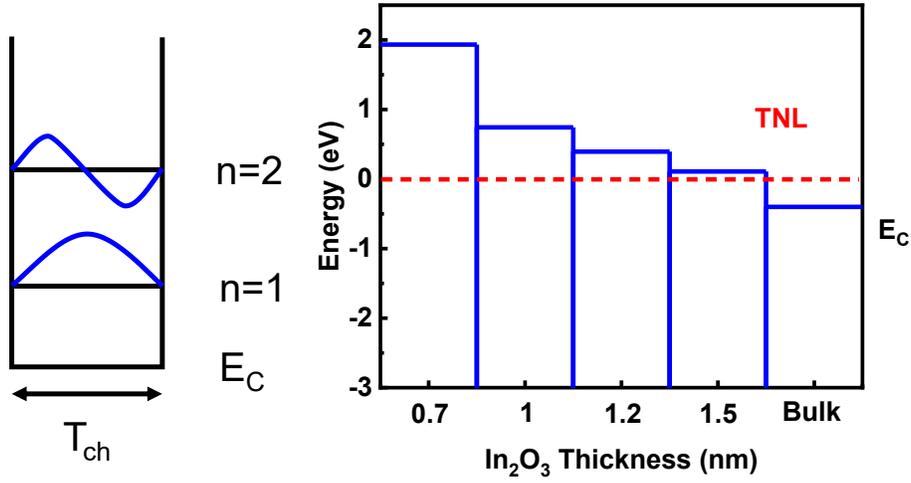

**Figure S3.** Conduction band minimum of $In_2O_3$ at different thicknesses from 0.7 nm to bulk. TNL alignments are modulated by the thickness of $In_2O_3$ due to quantum confinement.

In $In_2O_3$ transistor structure as shown in Fig 1(a), the semiconducting $In_2O_3$ is sandwiched by insulating $Al_2O_3$ and air, so that electron transport in $In_2O_3$ behaves like 2D electron gas in an infinity quantum well, as also shown in Fig. S3. Considering the ground state energy in an infinity quantum well, the change of $E_C$ by quantum confinement can be written as $\Delta E_C = \frac{\pi^2 \hbar^2}{2m^*} \frac{1}{T_{ch}^2}$. The effective mass (m*) of electron in $In_2O_3$ is about $0.33m_e$, where $m_e$ is electron rest mass.[1,2] Thus, thickness-dependent $\Delta E_C$ with respect to TNL can be calculated, as shown in Fig. S3. Since TNL is the intrinsic property of the material so that it is independent on the channel thickness. As can be seen, this simple model shows TNL moves deeper inside bandgap while decreasing the $T_{ch}$, resulting in the reduction of carrier density and positive $V_T$ shift, which agrees well with the experimental data and DFT calculation.